\documentclass[aps,pra,twocolumn,showpacs,superscriptaddress]{revtex4-1}

\usepackage{amsfonts,mathrsfs,amsmath,amsthm,amssymb}
\usepackage{epsfig,upgreek}
\usepackage{bm}
\usepackage{color}
\usepackage[dvipsnames]{xcolor}

\usepackage{diagbox}
\usepackage{siunitx}
\usepackage{mathtools}

\usepackage{rotating}
\usepackage{array,multirow,graphicx}
\usepackage{tabularx}

\newcolumntype{M}[1]{>{\centering\arraybackslash}m{#1}}

\usepackage{tabulary}
\newcolumntype{K}[1]{>{\centering\arraybackslash}m{#1}}
\usepackage{soul}
\usepackage{tikz}
\usepackage{xparse}

\NewDocumentCommand\UpArrow{O{1.0ex} O{black}}{%
   \mathrel{\tikz[baseline] \draw [->, line width=0.01pt, #2] (0,0) -- ++(0,#1);}
}

\makeatletter
\newsavebox{\@brx}
\newcommand{\llangle}[1][]{\savebox{\@brx}{\(\m@th{#1\langle}\)}%
 \mathopen{\copy\@brx\kern-0.5\wd\@brx\usebox{\@brx}}}
\newcommand{\rrangle}[1][]{\savebox{\@brx}{\(\m@th{#1\rangle}\)}%
 \mathclose{\copy\@brx\kern-0.5\wd\@brx\usebox{\@brx}}}
\makeatother

\begin{document}  
\title {\bf 
Variational preparation of entangled states 
on quantum computers
}

\author{Vu Tuan Hai}
\thanks{Electronic address: haivt@uit.edu.vn}
\affiliation{University of Information Technology, 
Ho Chi Minh City, 700000, Vietnam}
\affiliation{Vietnam National University, 
Ho Chi Minh City, 700000, Vietnam}

\author{Nguyen Tan Viet}
\affiliation{FPT University, Hanoi, Vietnam}

\author{Le Bin Ho}
\thanks{Electronic address: binho@fris.tohoku.ac.jp}
\affiliation{Frontier Research Institute 
for Interdisciplinary Sciences, 
Tohoku University, Sendai 980-8578, Japan}
\affiliation{Department of Applied Physics, 
Graduate School of Engineering, 
Tohoku University, 
Sendai 980-8579, Japan}

\date{\today}

\begin{abstract}
We propose a variational approach for preparing entangled quantum states on quantum computers. The methodology involves training a unitary operation to match with a target unitary using the Fubini-Study distance as a cost function. We employ various gradient-based optimization techniques to enhance performance, including Adam and quantum natural gradient. Our investigation showcases the versatility of different ansatzes featuring a hypergraph structure, enabling the preparation of diverse entanglement target states such as GHZ, W, and absolutely maximally entangled states. Remarkably, the circuit depth scales efficiently with the number of layers and does not depend on the number of qubits. Moreover, we explore the impacts of barren plateaus, readout noise, and error mitigation techniques on the proposed approach. Through our analysis, we demonstrate the effectiveness of the variational algorithm in maximizing the efficiency of quantum state preparation, leveraging low-depth quantum circuits.
\end{abstract}
%
%
\maketitle

\section{Introduction} 

Quantum computation leverages principles 
of quantum physics to perform calculations, 
and recent advances in engineering have led 
to the development of quantum computers
with great potential for practical applications
\cite{doi:10.1126/science.abb2823,
PRXQuantum.2.017001,
Ebadi2021,Pirandola2015,SPILLER200330,
365700, grover1996fast, 
PhysRevLett.103.150502,
PhysRevApplied.15.034068,
PhysRevA.101.010301}.
However, the current 
devices have limitations 
in qubit capacity and noise levels
(noisy intermediate-scale 
quantum devices or NISQ),
making them impractical 
for some applications
%
\cite{Preskill2018quantumcomputingin}. 

Despite this, various hybrid 
quantum-classical algorithms 
(VQAs) have been proposed 
and actively studied, 
showing promise for substantial 
quantum speedup in the NISQ regime 
\cite{Cerezo2021_r}. 
In a VQA scheme, 
a quantum circuit can be prepared, controlled, 
and measured in a quantum computer 
and then processed numerically 
in a classical computer. 
VQAs have been successfully 
applied to different problems such as  
variational quantum eigensolvers
\cite{Peruzzo2014,PhysRevResearch.1.033062,
Kirby2021contextualsubspace,Gard2020,PRXQuantum.2.020337},
quantum approximate optimization algorithms
\cite{PhysRevX.10.021067},
new frontiers in quantum foundations 
\cite{Arrasmith2019,
PhysRevLett.123.260505,
Koczor_2020,Meyer2021},
and so forth.

Recently, a quantum compilation scheme 
has gained significant attention 
due to its remarkable versatility 
in various applications
\cite{Jones2022robustquantum,PRXQuantum.2.040327,
Khatri2019quantumassisted,heya2018variational}. 
This technique involves training 
a trainable unitary to match it with
a target unitary \cite{heya2018variational,
Khatri2019quantumassisted}, 
which can be used for optimizing gates 
\cite{heya2018variational}, 
quantum-assisted compiling process
\cite{Khatri2019quantumassisted},
continuous-variable quantum learning
\cite{PRXQuantum.2.040327},
efficient quantum compilation
\cite{Jones2022robustquantum}, 
and quantum state tomography
\cite{Hai2023}.

In another aspect, 
quantum state preparation 
in quantum circuits has gained 
attention due to the advantages 
of quantum devices
\cite{aulicino2022state,
PhysRevResearch.4.013091,
Kuzmin2020variationalquantum,
lvovsky2009continuous, 
d2002quantum, 
takeda2021quantum}. 
Quantum computers allow for 
efficient preparation of quantum states, 
complete control of Hamiltonian 
for state evolution, and direct 
access to measurement results. 
Early works in this field can be divided 
into two categories: those with ancillary 
qubits and those without. 
Without ancillary qubits, 
the main challenge is the exponential growth 
of circuit depth, which can be as high as 
$\mathcal{O}[2^N]$ 
\cite{10.5555/2011670.2011675,
1629135,PhysRevA.93.032318,
PhysRevA.83.032302}, and
$\mathcal{O}[2^{N}/N]$
\cite{sun2021asymptotically}. 
Using ancillary 
qubits, on the other hand, 
significantly reduces the circuit depth 
to sub-exponential scalings, such as 
$\mathcal{O}[2^{N/2}]$ 
\cite{sun2021asymptotically,
PhysRevResearch.3.043200,
rosenthal2022query,Araujo2021}. 
However, this method still requires an 
exponential number of ancillary 
qubits in the worst cases.
Arrazola et al. recently used 
machine learning methods to effectively 
generate various photonic states,
which leverages an optimization method 
to find low-depth quantum circuits
\cite{Arrazola_2019}.

Despite progress being made,
preparing quantum states, 
especially entangled states, 
on NISQ devices is still a challenge. 
In this work, we propose a variational scheme 
bases on quantum compilation to 
prepare entangled quantum states. 
It utilizes a trainable unitary to act 
on a reference state (fixed as $|\bm 0\rangle$), 
which is then 
transformed into the desired target state. 
This method 
simplifies the preparation process and improves 
quantum circuit efficiency using trainable
unitaries with low depth.
Additionally, its adaptable nature 
enhances fault-tolerant capabilities.

Concretely, we first introduce the general 
framework of the variational 
quantum state preparation
using quantum compilation
and then apply it to prepare
entangled GHZ, W, and absolutely 
maximally entangled (AME) states. 
We propose several structures 
for the trainable unitary
using hypergraph-based ansatzes
and several gradient-based optimizers,
including the Adam and 
quantum natural gradient descent (QNG).
Here, we reduce the circuit depth 
from exponential to polynomial, 
i.e., $\mathcal{O}(L)$, 
$L$ the number of layers.
We also discuss the accuracy
under the effect of 
circuit depth, barren plateau, 
readout noise, 
and the error mitigation solution.
The results are applicable to 
any arbitrary target states. 

This paper is structured as follows.
Section~\ref{secii} describes the framework 
for variational quantum state preparation,  
including its ansatzes. 
Section ~\ref{seciii} showcases 
the numerical results, 
which are also thoroughly 
discussed in Sec.~\ref{seciv}. 
Finally, we conclude the work 
in Sec.~\ref{secv}.


\section{Framework}{\label{secii}}

\subsection{Variational quantum state preparation}
Quantum state preparation (QSP) 
is a process that uses 
controllable evolutions 
to transform the initial state 
of a quantum system into 
a desired target state.
This process is crucial for quantum 
computation and information processing. 
Here we present a variational scheme base 
on quantum compilation for the QSP 
with high accuracy and 
well against noise under mitigation aid.

\begin{figure}[t]
\includegraphics[width=8.6cm]{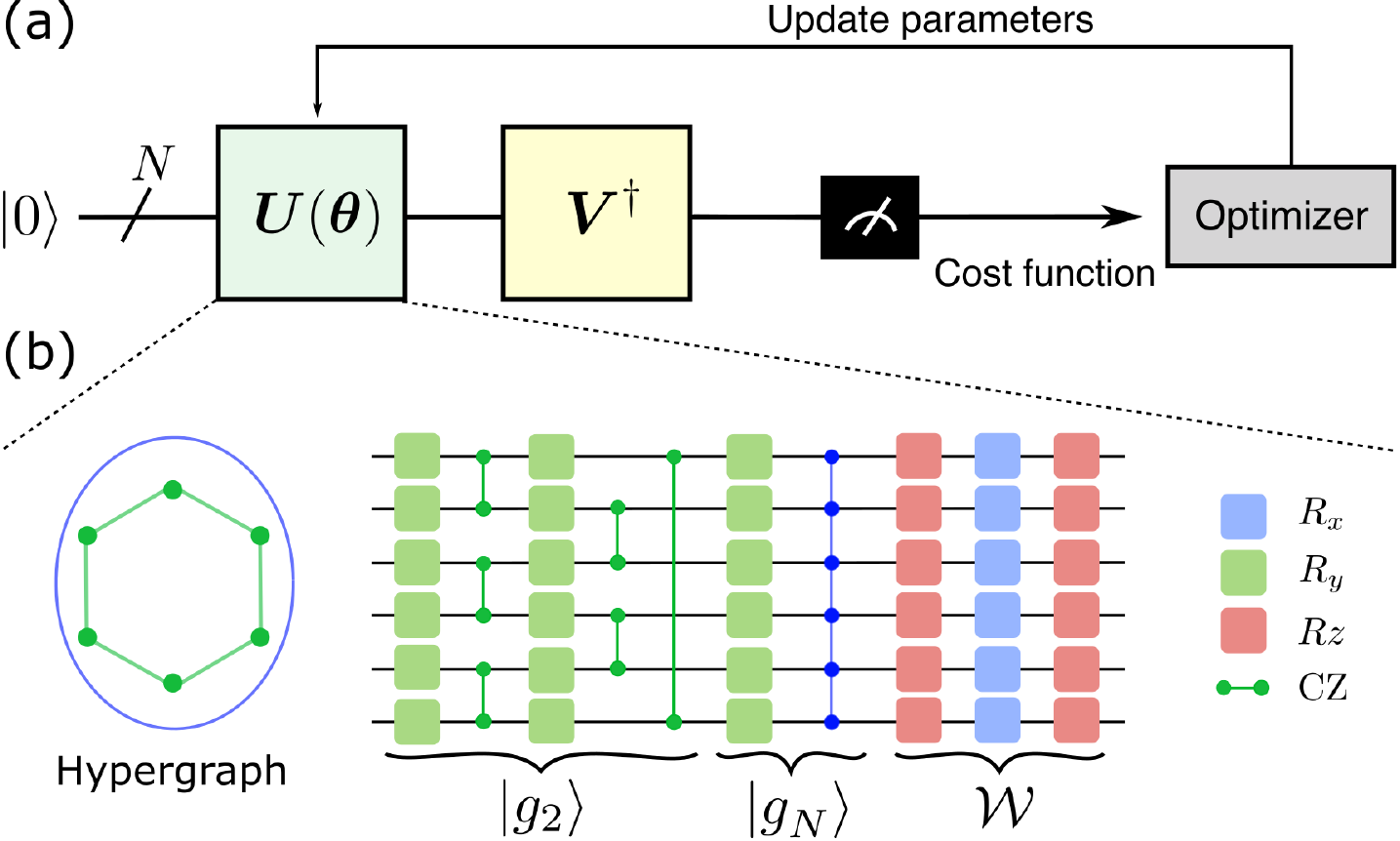}
\caption{
(a) Variational quantum algorithm (VQA) 
for quantum state preparation.
To begin, we
create the initial state
$|\bm 0\rangle$ and apply a series of 
$\bm U(\bm\theta)$ followed by 
$\bm V^\dagger$.
The resulting circuit is measured in
$|\bm 0\rangle\langle\bm0|$, 
and the outcome is sent to 
a classical computer for optimization
and update new parameters to the quantum circuit. 
(b) The ansatz $\bm U(\bm\theta)$ 
has a hypergraph-based structure,
which includes the rotation gates
$R_j, j = \{x, y, z\}$ and
the controlled-rotation-$Z$ gates (CZ).
}
\label{fig:1}
\end{figure}

Starting from the initial register 
$|\bm 0\rangle \equiv |0\rangle^{\otimes N}$
in the quantum circuit as shown in Fig.~\ref{fig:1}a, 
where $N$ is the number of qubits,
the task is to transform this state into 
a target quantum state, i.e., 
$|\psi\rangle$. 
We first transform the initial register 
$|\bm 0\rangle$
into a variational state 
\begin{align}\label{eq:psi_var}
|\phi(\bm\theta)\rangle = 
\bm U(\bm\theta)|\bm 0\rangle,
\end{align}
under a trainable unitary $\bm U(\bm\theta)$,
where $\bm\theta = \{\theta_1, 
\theta_2,\cdots, \theta_M\}$ can be 
adaptively updated during the training process, 
$M$ is the number of trainable parameters.
The target state can be expressed 
as 
\begin{align}\label{eq:psi_tar}
	|\psi\rangle = \bm V|\bm0\rangle,
\end{align}	
with a target (known) unitary $\bm V$.
The closeness of the two states 
is given by the Fubini-Study distance \cite{Kuzmak2021}
\begin{align}\label{eq:FB_dis}
	d(\phi(\bm\theta),\psi)
	= \sqrt{1-\big|\langle \psi |\phi(\bm\theta)\rangle\big|^2}
    = \sqrt{1-p_0(\bm \theta)},
\end{align}
where $p_0(\bm \theta) = 
|\langle \psi|\phi(\bm\theta)\rangle|^2
= |\langle \bm 0|
\bm V^\dagger\bm U(\bm\theta) |\bm 0\rangle|^2
$ is the probability 
of getting the outcome $|\bm 0\rangle$.
In practice, we start with the initial state 
$|\bm 0\rangle$ and apply a series of 
$\bm U(\bm\theta)$ followed by 
$\bm V^\dagger$  
to get the final state of 
$\bm V^\dagger\bm U(\bm\theta) |\bm 0\rangle$. 
We then  measure the projective operator 
$\bm P_0 = |\bm 0\rangle\langle\bm 0|$, 
which yields the probability $p_0(\bm\theta)$. 

The variational state becomes 
the target state when their distance reaches zero. 
We thus, use the Fubini-Study distance 
as a cost function, i.e.,
$\mathcal{C(\bm\theta}) = 
d(\phi(\bm\theta),\psi)$,
and minimize 
it 
\begin{align}\label{eq:optimize-theta}
    \bm\theta^*=\operatorname*{argmin}_{\bm\theta}
    \mathcal{C}(\bm{\theta}).
\end{align}
We train the variational circuit 
to reach the target state $|\psi\rangle$ 
represented by the corresponding
 $|\phi(\bm\theta^*)\rangle$.

\subsection {Hypergraph-based ansatzes}
A graph state is a configuration of multiple qubits 
that forms a graph structure 
with vertices and edges $\{V, E\}$. 
Each vertex represents a qubit and is connected to another 
qubit through an edge that uses a CZ gate for the interaction. 
In a regular graph state, 
every two qubits are connected in a pair:
\begin{align}\label{eq:graphstate}
|g_2\rangle = \prod_{\{i_1, i_2\}\in E}
{\rm CZ}_{i_1i_2}
|+\rangle,
\end{align}
where two vertexes $i_1, i_2$ 
connect via an edge,
and $|+\rangle = (|0\rangle + |1\rangle)/\sqrt{2}$.
A $k$-uniform hypergraph state is created 
in the same manner
by connecting $k$ vertices with an edge, 
as described in \cite{Rossi_2013}
\begin{align}\label{eq:hyper-graphstate}
|g_k\rangle = \prod_{\{i_1, i_2\cdots i_k\}\in E}
{\rm CZ}_{i_1i_2\cdots i_k}
|+\rangle.
\end{align}

Studies have shown that graph 
and hypergraph states are entangled states
\cite{PhysRevA.69.062311}. 
For example, a star graph is 
equivalent to a GHZ state 
with a local unitary (LU) operation
\cite{PhysRevA.69.062311,PhysRevLett.91.107903}, 
while a ring graph can be 
an AME state
\cite{PhysRevA.100.022342}.
These states have practical applications 
in quantum error correction
\cite{PhysRevA.105.042418,
PhysRevA.78.042303, 
hein2006entanglement}, 
quantum computing
\cite{PhysRevLett.86.5188,NIELSEN2006147},
quantum metrology
\cite{PhysRevLett.124.110502,le2023variational},
and other fields.

\begin{figure*}[t!]
\centering
\includegraphics[width=14.6cm]{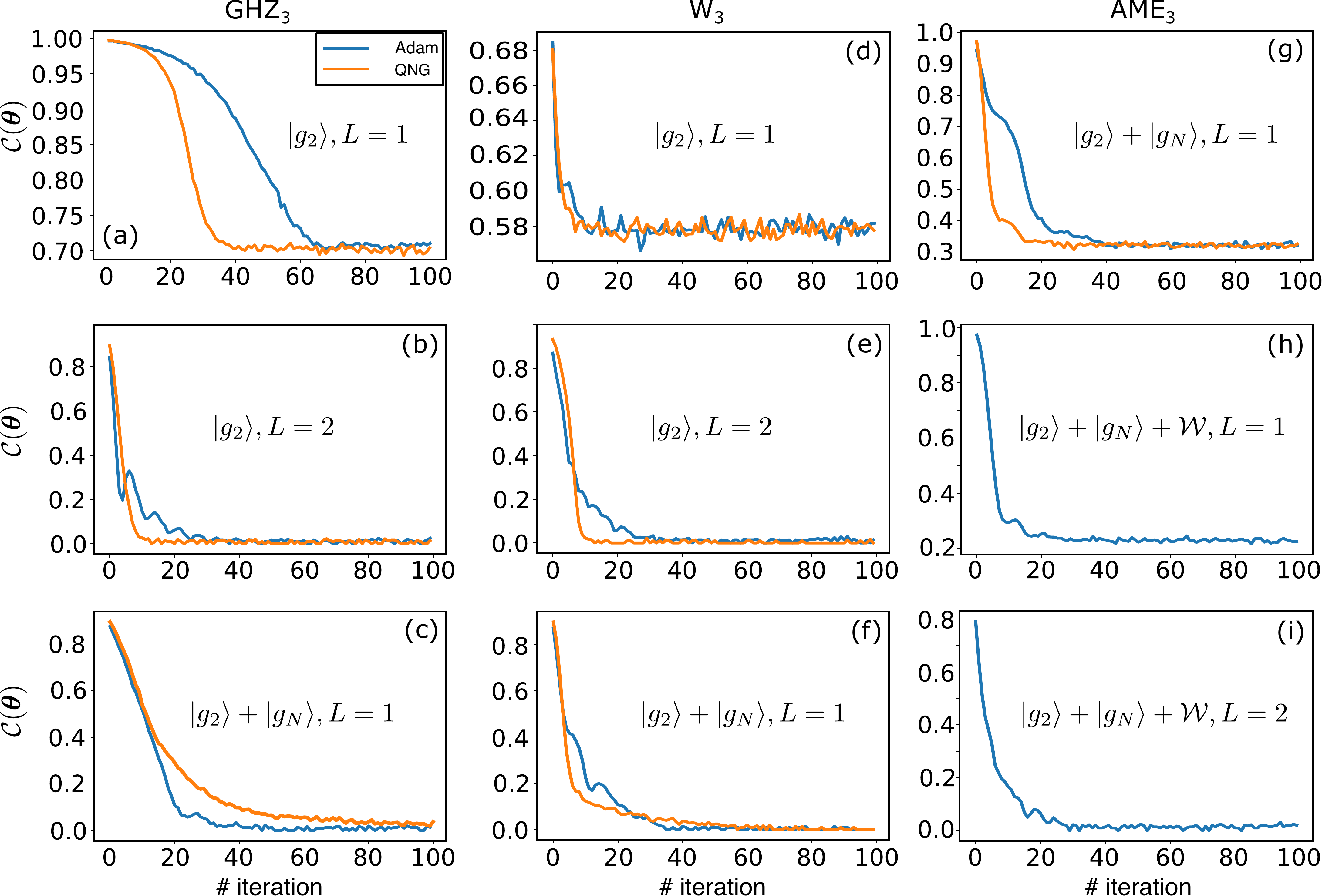}
\caption{
Cost function convergence for different 
target states and configuration ansatzes.
For the GHZ and W target states, the ansatz $|g_2\rangle, L = 1$ (a, d) 
fails to reach the global minimum, 
while increasing the ansatz depth to $L = 2$ (b, e) 
or incorporating $|g_N\rangle$ configuration (c, f) 
enables reaching the global minimum. 
For the AME target, the ansatz 
$|g_2\rangle + |g_N\rangle, L = 1$ (g) 
falls short of achieving the global minimum, 
but introducing the entangled term $\mathcal{W}$ 
improves the cost function (h) 
and leads to the global minimum (i). 
These results are shown for a system size of $N = 3$ qubits.
}
\label{fig:2}
\end{figure*}

This study uses hypergraph ansatzes 
to generate entangled states in quantum circuits. 
To create a graph-based ansatz, 
we use $R_y$ gates, with four $R_y$ 
gates surrounding each CZ gate. 
The formula for a single-qubit rotation is 
$R_y = \exp(-i\frac{\theta}{2}\sigma_y)$, with $\sigma_y$ 
representing a Pauli matrix. 
This structure is similar to the one in Ref.~\cite{Cerezo2021}.
We focus on combining 
a ring 2-uniform graph $|g_2\rangle$ 
and an $N$-uniform hypergraph $|g_N\rangle$, 
where $N$ corresponds to the number of qubits. 
We also combine it with an entangled  
$\mathcal{W}$ ansatz for ``phase" generation. 
For details, please refer to Fig.~\ref{fig:1}b. 
The circuit depth increases 
with the number of layers $L$, 
much smaller than sub-exponential. 
For a summary, please see Tab. ~\ref{tab:1}  
for a breakdown of these cases.
Here, the circuit depth is determined 
by calculating the longest path between 
the data input and the output.

\begin{table}[!t]
  \centering
    \caption{Hypergraph-based ansatzes and their characteristics.}
  \begin{tabular}{|c|c|c|c|}
    \hline
     ansatz & $N$ & 
     \# parameters &  
     circuit depth 
     \\\hline
     \multirow{2}{*}{$|g_2\rangle$} 
     &even& \multirow{2}{*}{$2NL$} & $4L$ \\ 
     \cline{2-2} \cline{4-4} 
     &odd& & $6L$ \\
     \hline
     \multirow{2}{*}{$|g_{2}\rangle + |g_{N}\rangle$}  
     &even& \multirow{2}{*}{$3NL$} & $6L$ \\
     \cline{2-2} \cline{4-4}
     &odd& & $8L$ \\
     \hline
     \multirow{2}{*}{$|g_{2}\rangle + |g_{N}\rangle + \mathcal{W}$}  
     &even& \multirow{2}{*}{$6NL$} & $9L$ \\
     \cline{2-2} \cline{4-4}   
     &odd& & $11L$ \\
     \hline
   \end{tabular}
  \label{tab:1}
\end{table}

\section{Numerical Results}\label{seciii}
We conduct numerical simulation 
to train the variational model. 
We use the Qiskit open-source package 
(version 0.39.4), which is compatible 
with all platforms, to execute the simulation. 
To obtain the probability $p_0$  for each experiment, 
we run $10^4$ shots using the {\it qasm} simulator backend. 
The training process involves 100 iterations.
Below we show the results of preparing some representative 
entangled states, including the GHZ, W, and AME states.

\subsection{Cost function: case studies}\label{seciii_i}

\begin{figure}[t]
\centering
\includegraphics[width=8.6cm]{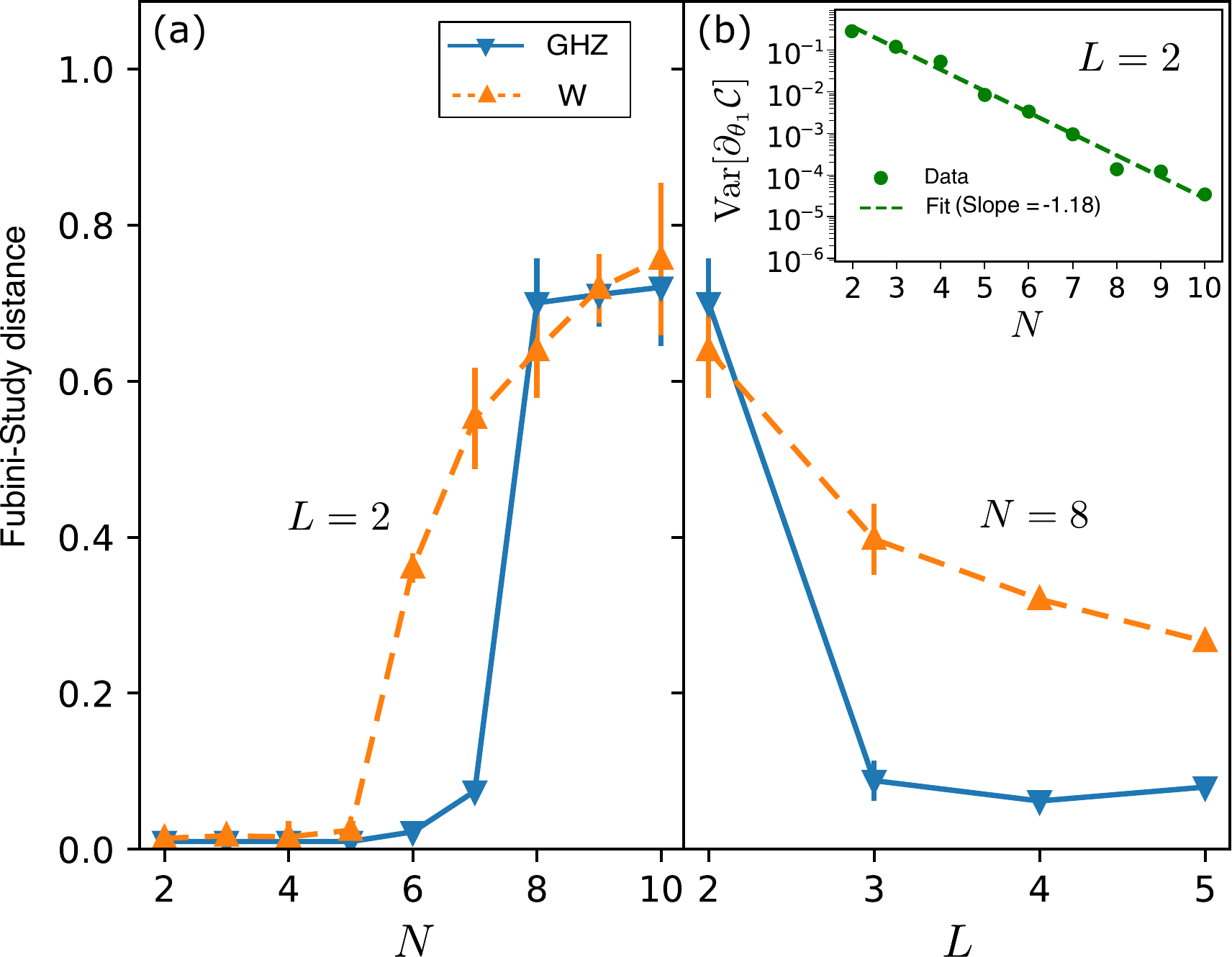}
\caption{
(a) Funibi-Study distance 
versus the number of qubits $N$ 
at $L = 2$.
We examine the target states GHZ and W. 
While the distance remains small up to $N = 5$,  
then it increases when increasing $N$.
(b) Fubini-Study distance versus 
the number of layers $L$ at $N = 8$.
For the GHZ target, 
the distance reaches the minimum
at $L = 4$, while for the W one, 
the distance continues to reduce.
(inset): A semi-log plot
of the variance
${\rm Var}[(\partial_{\theta_1}\mathcal{C})]$ 
as a function of $N$ at fixed $L = 2$ and the GHZ target. 
As predicted, exponential decay is observed 
as proof of the existence of the barren plateaus.
The slope of a fit line indicates 
the decay rate.
In all cases, we use 
$|g_2\rangle$ ansatz
and Adam optimizer.
}
\label{fig:3}
\end{figure}

In Fig.~\ref{fig:2}, we examine the cost function
versus the number of iterations for different configurations.  
We fix $N = 3$, and the target states are
\begin{align}\label{eq:3tar}
|\psi\rangle_{\rm GHZ} &= \dfrac{1}{\sqrt{2}}
(|000\rangle + |111\rangle),\\
|\psi\rangle_{\rm W} &= \dfrac{1}{\sqrt{3}}
(|001\rangle + |010\rangle + |100\rangle),\\
\notag |\psi\rangle_{\rm AME} &= 
0.27|000\rangle
+ 0.377 |100\rangle
+ 0.326 |010\rangle \\
&+ 0.363 |001\rangle
+ 0.74e^{-0.79i\pi}|111\rangle,
\end{align}
where the AME state is taken from 
Ref.~\cite{Enriquez2016}.

Figure \ref{fig:2}(a-c) display results for the GHZ case.
When using the ansatz $|g_2\rangle$ with $L = 1$, 
the cost function $\mathcal{C}(\bm\theta)$ 
reaches a local minimum of 
approximately 0.7 (Fig.~\ref{fig:2}a)
due to the limited parameters space. 
However, at $L = 2$, $\mathcal{C}(\bm\theta)$ 
reaches the global minimum of zero (Fig.~\ref{fig:2}b).
Similar outcomes are observed 
when using $|g_2\rangle + |g_N\rangle$ 
and maintaining $L = 1$, 
as shown in Figure~\ref{fig:2}c.
The results for the W case, 
as shown in Figs.~\ref{fig:2}(d-f), 
share similarities 
with the outcomes of the GHZ case.
In the AME case, 
achieving the global minimum with 
$|g_2\rangle + |g_N\rangle, L = 1$ 
configuration (Fig.~\ref{fig:2}g) 
is challenging because the AME 
state requires a higher degree of entanglement.
To enhance the optimal outcomes, 
we incorporate an entangled term 
$\mathcal{W}$ into the hypergraph-based ansatz.
The configuration $|g_2\rangle + |g_N\rangle + \mathcal{W}$ 
yields $\mathcal{C}(\bm\theta) \approx 0.2$ 
with $L = 1$ (Fig.~\ref{fig:2}h) 
and $\mathcal{C}(\bm\theta) \approx 0.0$ 
with $L = 2$ (Fig.~\ref{fig:2}i). 
Hence, the cost functions for all case studies 
presented here can attain 
the global minimum at a specific configuration.
Upon comparing the Adam and QNG optimizers, 
it appears that the QNG optimizer 
reaches convergence in fewer iterations 
due to its superior ability to 
navigate toward the optimal direction 
\cite{Stokes2020quantumnatural,haug2021natural}.

\subsection{Funibi-Study distance vs $N$}
In Fig. \ref{fig:3}a, 
we present the numerical results 
for the Fubini-Study distance as a 
function of the number of qubits $N$. 
Specifically, we focus on 
$|g_2\rangle$ configuration with $ L = 2$, 
averaging the results over 10 repeated experiments 
using the Adam optimizer. 
Generally, the Fubini-Study distance 
tends to increase with the number of qubits, 
indicating lower accuracy in larger systems. 
However, in this analysis, we demonstrate that 
the Fubini-Study distance remains small, 
even up to $N = 5$, suggesting high accuracy 
and stability within this range. 
Beyond $N = 5$, the Fubini-Study distance 
gradually increases as $N$ increases. 
Moreover, preparing the target W state 
presents more significant challenges 
as it belongs to the multipartite 
entanglement class
\footnote{For W states, if one qubit is lost, 
the remaining system still entangles,
from which contrasts with GHZ states, 
that fully separable after disentangling one qubit.
See also Ref.~\cite{PhysRevA.62.062314}}.

\subsection{Funibi-Study distance vs $L$}
The accuracy of the training process 
is significantly affected by the number of layers $L$
\cite{Steinbrecher2019,Giacomo2020,
PhysRevA.104.042601}. 
In Fig.~\ref{fig:3}b we demonstrate 
the correlation between the Fubini-Study distance 
and the number of layers using 
$|g_2\rangle$ ansatz with $N = 8$ 
and the Adam optimizer. The Fubini-Study distance 
reaches its minimum at $L = 4$ for the GHZ target 
and slightly increases when $L > 4$.
This observation aligns with the barren plateaus 
analyses depicted in the inset of Fig.~\ref{fig:3} below. 
Concerning the W state, 
the distance decreases until $L = 5$. 
This finding confirms the previous observation 
regarding the entangled characteristics 
of the GHZ and W classes, where achieving 
the same level of accuracy as the GHZ 
class requires a greater computational 
cost for preparing the W class. 

\subsection{Barren plateau}
The results above indicate the presence 
of barren plateau effect (BP) 
in the training landscapes \cite{McClean2018}, 
a phenomenon commonly observed in 
variational quantum algorithms 
and quantum neural networks.
When the BP occurs, 
the derivative of the cost function 
$\mathcal{C}(\bm\theta)$ exponentially 
diminishes with increasing 
circuit size \cite{McClean2018}. 
The BP effect can arise from multiple factors, 
including random parameter initialization \cite{McClean2018}, 
shallow depth with global cost functions \cite{Cerezo2021}, 
highly expressive ansatzes \cite{PRXQuantum.3.010313}, 
entanglement-induced effects \cite{PRXQuantum.2.040316}, 
and noise-induced influences \cite{Wang2021}.
Numerous strategies have been 
identified to bypass the BP effect, 
such as employing local cost functions 
\cite{Cerezo2021,Uvarov_2021}, 
utilizing correlated parameters 
\cite{Volkoff_2021}, 
pretraining with classical neural networks 
\cite{verdon2019learning}, 
and adopting layer-by-layer training approaches 
\cite{grant2019initialization}.

To investigate the disappearance of 
the derivative cost function 
$\mathcal{C}(\bm\theta)$ 
and the presence of BP, 
we calculate the variance 
\begin{align}\label{eq:varC}
{\rm Var}[\partial_k\mathcal{C}] = 
\langle(\partial_k\mathcal{C})^2\rangle
-\langle\partial_k\mathcal{C}\rangle^2,
\end{align}
where we used $\mathcal{C}$ 
to represent $\mathcal{C}(\bm\theta)$,  
$\partial_k\mathcal{C}$ to represent 
$\partial\mathcal{C}/\partial\theta_k$,
and the expectation value is taken over the final state. 
The numerical result ${\rm Var}[\partial_{\theta_1}\mathcal{C}]$ 
for a representative first parameter $\theta_1$ 
is given in Fig.~\ref{fig:3} inset. 
It demonstrates an exponential decay 
with a slope of -1.18 as the number of qubits $N$ increases. 
This is evidence for the existence of BP 
within the parameters space in our model 
and agrees with the prediction in main Fig.~\ref{fig:3}. 
The result is shown for the GHZ target and $L = 2$.

\subsection{Error mitigation}
Finally, we consider the impact of noise 
and employ error mitigation techniques. 
In the current era of noisy 
intermediate-scale quantum computers (NISQ), 
the influence of noisy qubits induces biases 
and thus restricts the range of 
applications \cite{Nachman2020}. 
One of the important classes of 
noisy qubits is the readout error, 
which typically arises from (i) 
qubits decoherence, i.e., qubits decay, 
phase change,..., during the measurement time, 
and (ii) incomplete measuring devices, i.e., 
the overlap between measured bases.
Here we model the noise channel by 
a readout error probability for 
each qubit in the circuit as
\begin{align}\label{eq:prob}
	\begin{pmatrix}
		p^{(\epsilon)}_0\\ p^{(\epsilon)}_1
	\end{pmatrix}
	=
	\begin{pmatrix}
	1-\epsilon & \epsilon\\
	\epsilon & 1-\epsilon
	\end{pmatrix}	
	\begin{pmatrix}
		p_0\\ p_1
	\end{pmatrix},
\end{align}
where $p_0, p_1$ are the true 
probabilities when measuring
two elements $|0\rangle$ and $|1\rangle$
of the qubit, and $p^{(\epsilon)}_0, 
p^{(\epsilon)}_1$ are the 
readout error probabilities, respectively,
$\epsilon$ is the error rate.

\begin{figure} [t!]
\includegraphics[width=8.6cm]{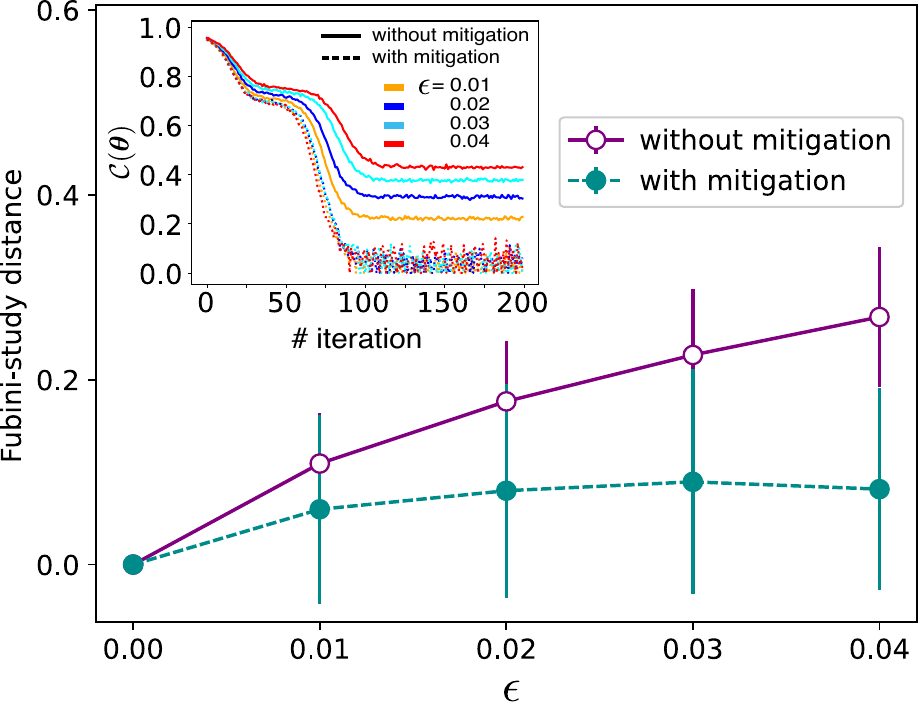}
\caption{Fubini-Study distance 
under the presence of noise 
($\epsilon$)
with and without mitigation. 
Inset: plot of cost function versus iteration
for two cases of with and without mitigation 
with various error rates $\epsilon$.
We fixed $N = 5, L = 2$, QNG optimizer, 
and plot for $|g_2\rangle$ ansatz with 
the GHZ target.
}
\label{fig:4}
\end{figure}

In Fig.~\ref{fig:4} (open circle, \textcolor{Plum}{$\circ$}),
we observe a rapid increase in the 
Fubini-Study distance as the error rate rises. 
The inset of Fig.~\ref{fig:4} displays 
the cost function for different 
$\epsilon$ values (the dotted lines), 
with rapid reduction and convergence 
around 100 iterations. However, for $\epsilon \in [0.01, 0.04]$, 
the optimal values are limited to $[0.22, 0.42]$. 
This results in an increased 
Fubini-Study distance in the main figure, 
leading to a loss of accuracy in the presence of noise.

To suppress the noise, 
we apply the measurement error mitigation technique 
\cite{doi:10.7566/JPSJ.90.032001,
Czarnik2021errormitigation,
PRXQuantum.2.040326,
Maciejewski2020mitigationofreadout}.
The basic ideal behind this method 
is to utilize a calibration matrix $\mathcal{M}$ 
of size $(2^N\times2^N)$, such that
\begin{align}\label{eq:M}
	\bm p^{(\epsilon)} =  \mathcal{M}\bm p,
\end{align}
where vector $\bm p = (p_0, p_1, 
\cdots, p_{2^N-1})^{\rm T}$ 
represents the true probabilities, 
and similar for $\bm p^{(\epsilon)}$ 
which represents 
the readout probabilities.
The calibration matrix comprises 
all the transition probabilities, 
denoted a $\mathcal{M}_{ij} = {\rm Prob}
(p^{(\epsilon)}_i\to p_j)$.
An effective method 
to obtain the mitigated 
probabilities is inverting 
the calibration matrix, 
resulting in
$\bm p^{\rm mitigated} = \mathcal{M}^{-1}
\bm p^{(\epsilon)}$.
Various techniques have been demonstrated, 
including the least square \cite{Geller_2020}, 
truncated Neumann series \cite{wang2021measurement}, 
and unfolding methods \cite{Nachman2020}.
Here we limit ourselves to the direct approach 
which suffices for our analysis.
We construct $\mathcal{M}$ 
by running $2^N$ circuits corresponding to  
$2^N$ elements in the computation basis 
$\{|00\cdots0\rangle, |00\cdots1\rangle, 
\cdots, |11\cdots1\rangle\}$.

The mitigation results are depicted in 
Fig.~\ref{fig:4} (filled circle, 
\textcolor{JungleGreen}{$\bullet$}). 
In this specific instance, 
noise is effectively mitigated, 
leading to a significant reduction 
in the Fubini-Study distance, 
which becomes comparable to 
its original values before 
introducing noise. The average results 
are graphed after conducting 10 
repeated experiments 
using the QNG optimizer.

\begin{figure*} [t!]
\includegraphics[width=12.60cm]{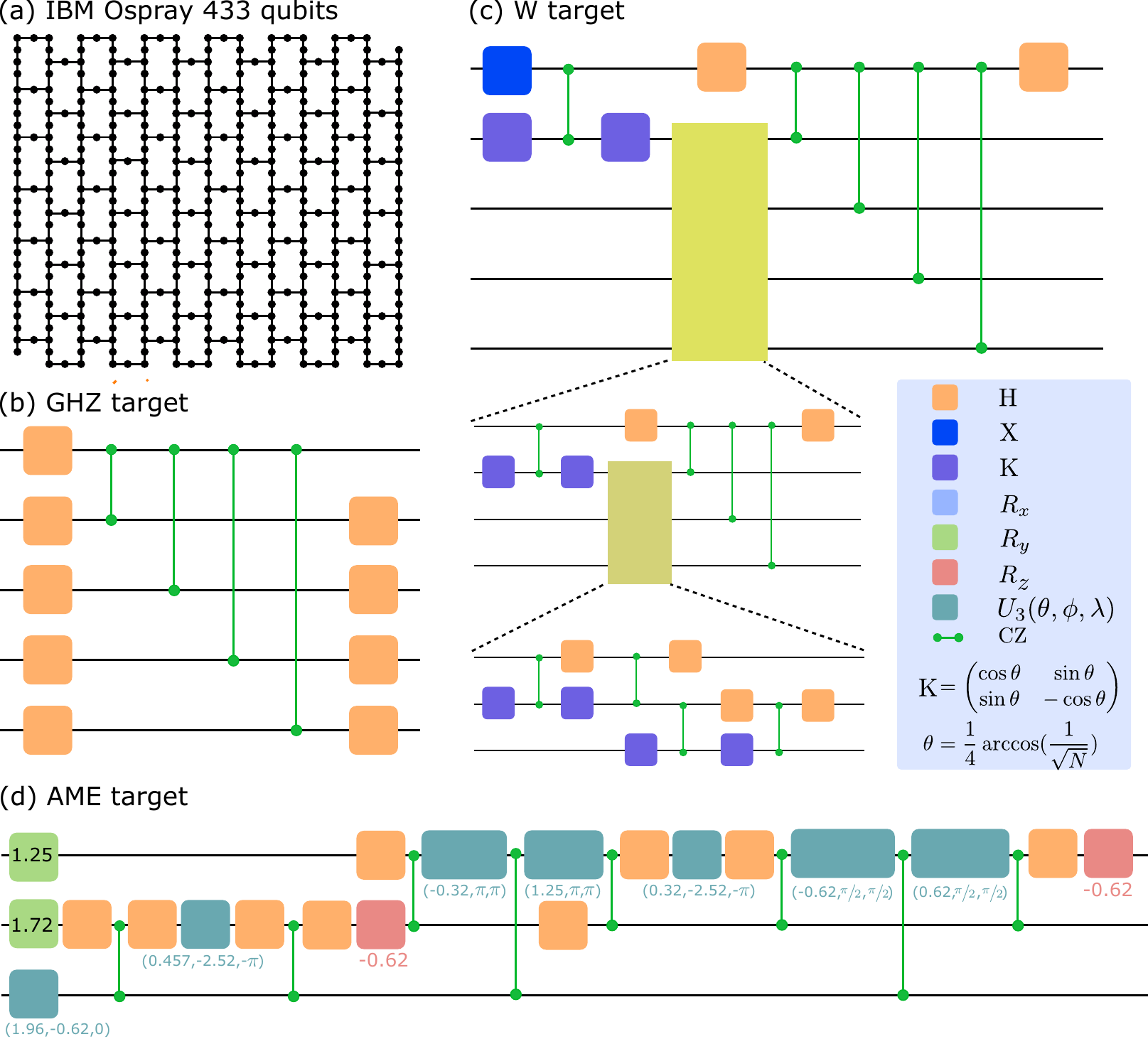}
\caption{(a) The structure of 
IBM Osprey with 433 qubits, 
represented by dots, 
and the lines indicate 
their physical interaction. 
(b) The quantum circuit prepares 
the target GHZ state, 
with an example of $N = 5$. 
(c) The quantum circuit prepares 
the target W state, 
with an example of $N = 5$. 
(d) The quantum circuit prepares 
the target AME state, 
with an example of $N = 3$.
}
\label{fig:5}
\end{figure*}

\section{Discussion}\label{seciv}
We emphasize two crucial advantages of the framework. 
Firstly, graph-based ansatzes exhibit 
high versatility and are compatible 
with the current superconductor-based 
quantum hardware architectures. 
For example, the IBM Osprey system, 
boasting 433 qubits, features a ring structure
as depicted in Fig.~\ref{fig:5}a, 
where the nodes represent qubits, 
and the lines show their physical connections. 
In our approach, we can design graph-based 
ansatzes that align with this hardware's 
vertices and edges, allowing for 
the utilization of various configurations 
such as linear, star, ring, 
and hypergraph ansatzes. 
This adaptability of variational ansatzes 
proves instrumental in effectively 
mitigating physical noise 
inherent in the system.

To prepare a GHZ state in quantum circuits, 
there are two approaches available. 
The first method directly utilizes 
the target unitary, as illustrated in 
Fig.~\ref{fig:5}b. 
However, this approach necessitates 
the interaction of at least one qubit 
with the remaining $N - 1$ qubits, 
thereby restricting the preparation 
to cases where $N\le 4$. 
On the other hand, employing a trainable ansatz 
with a ring structure can achieve 
comparable accuracy for any value of $N$, 
as illustrated in Fig.~\ref{fig:3}. 
However, as the value of $N$ increases, 
the worst-case scenarios may require 
a more significant number of layers, 
resulting in a deeper circuit configuration.

Another crucial aspect of our framework 
is the ability to reduce the circuit depth 
for trainable ansatzes. With this approach, 
the circuit depth remains constant 
regardless of the number of qubits $N$, 
enabling scalability. We have effectively 
demonstrated this scalability in the 
cases of W and AME, where the circuit depth 
of the trainable ansatzes is shorter 
than their respective target 
configurations (refer to Fig.~\ref{fig:5}(c, d) 
for the target unitaries of the W and AME states). 
Table~\ref{tab:2} summarizes the circuit depth 
for $N$ up to 5, explicitly comparing cases 
where the trainable ansatzes successfully 
recover the target unitary, indecated 
by $\mathcal{C}(\bm\theta) = 0$.

\begin{widetext}
\begin{table*}[t]
\centering
\caption{Comparison of circuit depth 
between target and trainable unitaries 
for $N$ ranging from 2 to 5.}
\begin{tabular}{|c|c|c|c|c|c|c|c|c|c|c|c|c|}
   \hline
   $N$ & 
   \multicolumn{3}{c|}{2} & 
   \multicolumn{3}{c|}{3} & 
   \multicolumn{3}{c|}{4} & 
   \multicolumn{3}{c|}{5} \\
   \hline
   \diagbox[]{ansatzes}{targets} & 
   GHZ & 
   W & 
   AME &
   GHZ & 
   W & 
   AME & 
   GHZ & 
   W & 
   AME &
   GHZ & 
   W &
   AME\\
   \hline
   $|g_2\rangle, L = 2$ & 
   \diagbox[]{8}{3} & 
   \diagbox[]{8}{2} & 
   - &
   \diagbox[]{12}{4} & 
   \diagbox[]{12}{8} & 
   - & 
   \diagbox[]{8}{5} & 
   \diagbox[]{8}{15} & 
   - &
   \diagbox[]{12}{5}& 
   \diagbox[]{12}{23} &
   -
   \\
   \hline
   $|g_2\rangle + |g_N\rangle, L = 1$ & 
   \diagbox[]{6}{3} & 
   \diagbox[]{6}{2} & 
   - &
   \diagbox[]{8}{4} & 
   \diagbox[]{8}{8} & 
   - & 
   \diagbox[]{6}{4} & 
   \diagbox[]{6}{15} & 
   - &
   \diagbox[]{8}{5}& 
   \diagbox[]{6}{23}&
   -
   \\
   \hline
   $|g_2\rangle + |g_N\rangle + \mathcal{W}, L = 2$ & 
   - & 
   - & 
   - &
   - & 
   - & 
   \diagbox[]{22}{24} & 
   - & 
   - &
   - &
   -& 
   - &
   -
   \\
   \hline
\end{tabular}
\label{tab:2}
\end{table*}
\end{widetext}

\section{Conclusion}\label{secv}
We presented a versatile variational algorithm 
for quantum state preparation 
that explicitly compiles a given 
quantum state to another. We thoroughly 
analyzed the algorithm's performance 
through extensive numerical experiments 
using various hypergraph-based ansatzes 
and optimizers. Our algorithm offers the 
advantage of creating low-depth circuits, 
making it highly suitable for implementation 
on different quantum hardware platforms. 
Notably, it can be utilized for testing 
the capabilities of quantum hardware 
in various applications 
\cite{PhysRevA.100.022342}.

Furthermore, we demonstrated the efficient 
preparation of entanglement classes and 
the effectiveness of employing an error 
mitigation protocol to suppress noise. 
This capability extends to any desired 
target state. Additionally, we explored 
the impact of the barren plateau phenomenon, 
an inherent consequence of expanding the system space.


\begin{acknowledgments}
This work was supported by JSPS KAKENHI Grant
Number 23K13025 
and the VNUHCM-University of 
Information Technology’s 
Scientific Research Support Fund.
\end{acknowledgments}
\vspace{0.25cm}
\noindent {\bf Code availability}:
The codes used for this study are available 
in https://github.com/vutuanhai237/UC-VQA.


\appendix
\setcounter{equation}{0}
\renewcommand{\theequation}{A.\arabic{equation}}

\setcounter{equation}{0}
\renewcommand{\theequation}{B.\arabic{equation}}

\bibliography{refs}
\end{document}